# Influence of electron irradiation on fluctuation conductivity and pseudogap in YBa$_2$Cu$_3$O$_{7-\delta}$ single crystals

A. L. Solovjov[1,2,3], L. V. Omelchenko[1], E. V. Petrenko[1,4], G. Ya. Khadzhai[2],
D. M. Sergeyev[4], A. Chroneos[5,6], and R. V. Vovk[2]

*B. Verkin Institute for Low Temperature Physics and Engineering of the National Academy of Sciences of Ukraine*
*Kharkiv 61103, Ukraine*
E-mail: solovjov@ilt.kharkov.ua

[2]*Department of Physics, V. N. Karazin Kharkiv National University, Kharkiv 61022, Ukraine*

[3]*Institute for Low Temperatures and Structure Research, Polish Academy of Sciences, Wroclaw 50-422, Poland*

[4]*K. Zhubanov Aktobe Regional State University, Aktobe 030000, Kazakhstan*

[5]*Department of Materials, Imperial College, London SW7 2AZ, UK*

[6]*Department of Electrical and Computer Engineering, University of Thessaly, Volos 38221, Greece*

The effect of electron irradiation with the energy of 2.5 MeV on the temperature dependences of the resistivity $\rho(T)$ of an optimally doped YBa$_2$Cu$_3$O$_{7-\delta}$ single crystal has been studied. The temperature dependences of both fluctuation conductivity $\sigma'(T)$ and the pseudogap $\Delta^*(T)$ on irradiation dose $\varphi$ have been calculated within the local pair model. Here we show that with an increase in $\varphi$, the value of $\rho(300\text{ K})$ increases linearly, while $T_c$ decreases linearly. Concurrently, the value of $\rho(100\text{ K})$ increases nonlinearly, demonstrating a feature for $\varphi_3 = 4.3\cdot10^{18}$ e/cm$^2$, which is also observed in the number of other dose-dependent parameters. Regardless of the irradiation dose, in the temperature range from $T_c$ up to $T_{01}$, $\sigma'(T)$ obeys the classical fluctuation theories of Aslamazov–Larkin (3D–AL) and Maki–Thompson (2D–MT), demonstrating 3D–2D crossover with increasing temperature. The crossover temperature $T_0$ makes it possible to determine the coherence length along the $c$ axis, $\xi_c(0)$, which increases by ~ 3 times under irradiation. Furthermore, the range of superconducting fluctuations above $T_c$ also noticeably increases. At $\varphi_1 = 0$, the dependence $\Delta^*(T)$ typical for single crystals containing pronounced twin boundaries is observed with a maximum at $T_{pair} \sim 120$ K and a distinct minimum at $T = T_{01}$. It was determined for the first time that at $\varphi_3 = 4.3\cdot10^{18}$ e/cm$^2$ the shape of $\Delta^*(T)$ changes strongly and becomes the same as in optimally doped YBa$_2$Cu$_3$O$_{7-\delta}$ single crystals with a very low pseudogap opening temperature $T^*$ and noticeably reduced $T_{pair}$, while at $T_c(\varphi)$ there are no singularities. With an increase in the irradiation dose up to $\varphi_4 = 8.8\cdot10^{18}$ e/cm$^2$, the shape of $\Delta^*(T)$ is restored and becomes the same as in well-structured YBa$_2$Cu$_3$O$_{7-\delta}$ films and untwined single crystals. Moreover, in this case, $T_{pair}$ and $T^*$ increase noticeably.

## 1. Introduction

Elucidation of the conditions for the occurrence of high-temperature superconductivity and the study of the physical properties of high-temperature superconductors (HTSCs) continues to be one of the most important areas of solid-state physics. Despite the fact that more than 35 years have passed since the discovery of HTSCs, its microscopic mechanism is still controversial. According to modern concepts, the key to understanding the nature of the superconducting state and various physical properties of HTSCs materials is the study of unusual phenomena that manifest themselves in these compounds in the normal (i.e., non-superconducting) state. These phenomena, in particular, include the anomalously wide temperature range of the excess conductivity manifestation, the so-called

"pseudogap" (PG) [1–5], incoherent electrical transport [6], metal–insulator transitions [7, 8], Lifshitz transition [9], etc.

A key issue of HTSCs physics is the study of the anomalies of the pseudogap state [1–5, 9–15], which is observed below the PG opening temperature, $T^* \gg T_c$, in the range of the phase diagram with a concentration of charge carriers less than the optimal one, which is usually called the range of "underdoped states". In this region, numerous anomalies of electronic properties are observed, associated with a decrease in the density of single-particle excitations and anisotropic rearrangement of the spectral density of charge carriers, most likely due to the rearrangement of the Fermi surface [9, 10, 16]. Understanding the nature and properties of the PG state is a central problem in any approach to describe the complex nature of HTSC systems.

Despite a very large number of experimental and theoretical works devoted to the study of these phenomena, there is still no consensus regarding the understanding of the mechanisms leading to their occurrence. This is largely due to the fact that a significant part of the experimental data was obtained on ceramic, textured polycrystalline and film samples. In this regard, special relevance is addressed to the study of physical properties of HTSCs on single-crystal samples with a given topology of the defect structure, as well as using experimental techniques using extreme conditions (low temperatures [6], high values of the magnetic field [9, 10], and high pressures [17]). These can create additional defects in HTSC samples, the role of which is also not fully understood (refer to [17] and references therein).

Various structural defects significantly affect the behavior of HTSCs, increasing the resistivity ρ, but decreasing the superconducting transition temperature $T_c$ and increasing its width $\delta T_c$ [17–20]. As a rule, single crystals of HTSCs (cuprates), such as REBa$_2$Cu$_3$O$_{7-\delta}$ (RE = Y and other lanthanides) without impurities, always contain defects in the form of twinning planes, which arise during the "tetra-ortho transition" and minimize the elastic energy of the crystal [21]. Such defects form parallel twin boundaries (TBs), which are planes with a tetragonal structure as a result of the presence of layers containing oxygen vacancies located along the TBs [22, 23]. Thus, TBs are regions of the sample with a reduced density of charge carriers and, as a consequence, with a lower $T_c$ compared to the rest of the array of single crystal. In this case, the various properties of such single crystals, measured in the experiment, will substantially depend on how the transport current flows: parallel [24] or at some angle to the TBs [17, 21]. Point defects (oxygen vacancies) are also present in all YBa$_2$Cu$_3$O$_{7-\delta}$ samples, which is associated with the non-stoichiometric oxygen content. Apart from the aforementioned defects, additional defects may occur in HTSCs, for example, by doping cuprates [usually, YBa$_2$Cu$_3$O$_{7-\delta}$ (YBCO)] with praseodymium [25, 26] which, like other RE, replaces Y. It is believed that such defects prevent the formation of local pairs (LPs), which, according to the model of local pairs, are responsible for the formation of the PG state in HTSCs [5, 12, 27, 28].

Of particular interest are the defects that arise when HTSCs are exposed to various irradiation (see [18] and references therein), and especially when irradiated with high-energy electrons [18–20, 29–33], since HTSC devices often have to operate under conditions of such irradiation [20, 30]. Electron irradiation is an appropriate method to study the effect of certain lattice defects on superconducting (SC) properties because by varying the electron energy we can selectively produce defects on the different sublattices. Thus, electron irradiation is a controlled and highly effective way to create a significant number of defects without changing the composition of the irradiated sample. In other words, electron irradiation allows a controlled change in the degree of disorder in the sample, gradually increasing the irradiation dose [20, 31, 33, 34]. On the other hand, it can be also expected that radiation defects will lead to pair-breaking of fluctuating Cooper pairs above $T_c$, thus affecting the fluctuation conductivity (FLC) and PG parameters. However, as far as we know, the study of the effect of electron irradiation on the FLC and PG in YBCO has not been carried out to date. The purpose of this work is to bridge this gap.

In this paper, we report on the study of the 2.5 MeV electron irradiation on the temperature dependences of the resistivity ρ(T), fluctuation conductivity, σ′(T), and pseudogap, Δ*(T), of optimally doped single crystal of YBa$_2$Cu$_3$O$_{7-\delta}$ with an increase in the irradiation dose from φ$_1$ = 0 up to φ$_4$ = 8.8·10$^{18}$ e/cm$^2$. The studies were carried out in a wide range of temperatures, from room temperature down to $T_c$, which may be the key to understanding the HTSCs nature and provide important information on the interaction of charge carriers with the phonon and defect subsystems. It is shown that irradiation monotonously increases ρ and ξ$_c$(0), and also decreases $T_c$ and Δ*($T_G$). However, the σ′(T) and Δ*(T) shape changes non-monotonically, demonstrating a bright feature at φ$_3$ = 4.3·10$^{18}$ e/cm$^2$, which is absent in the temperature dependences of a number of other measured parameters. A detailed discussion of the results obtained is given below.

## 2. Experiment

YBa$_2$Cu$_3$O$_{7-\delta}$ single crystals were grown by the solution-melt technology in gold crucibles, as described elsewhere [35]. The crystals were small-sized platelets with typical dimensions of (1.5...2)×(0.2...0.3)×(0.01...0.02) mm, where the smallest size corresponded to the $c$ axis. The small size of the samples was necessary to minimize the time of irradiation with a limited irradiating current. Selected crystals were annealed in oxygen at 650 °C for 1 h and then at 450 °C for several days [35, 36]. Annealing is used to obtain samples with the appropriate oxygen content [35–37] and transition temperatures $T_c$ of the annealed crystals in the range 90–92 K with a transition width $\delta T_c$ of < 1 K. Importantly, as mentioned above, the main part of our crystals contained defects in the form of pronounced TBs. It was expected that exactly the interaction of TBs with radiation defects will greatly affect the single crystal characteristics during irradiation.

A fully computerized setup utilizing the four-point probe technique with a stabilized measuring current of up to 10 mA was used to measure the *ab* plane resistivity, $\rho_{ab}(T)$. Silver epoxy contacts were glued to the opposite ends of the crystal in order to produce a uniform current distribution in the central region where voltage probes in the form of parallel stripes were placed with a distance of ~ 1 mm between them. The contact resistances did not exceed 1 $\Omega$. Temperature measurements were performed with a Pt sensor with an accuracy of about 1 mK [21].

The irradiations were performed with 2.5 MeV electrons in the low *T* facility of the Van de Graaff accelerator at the National Science Center "Kharkiv Institute of Physics and Technology" (Kharkiv, Ukraine). The samples were immersed in liquid He and the electron flux was limited to $10^{14}$ (e/cm$^2$)/s to avoid heating of the samples [30]. The sample thicknesses (≈ 10–60 µm) were much smaller than the penetration depth of the electrons, which ensured a homogeneous distribution of induced defects over the sample volume. The calculated energy losses for 2.5 MeV electrons for YBCO are ~ 1 keV/µm (see [20, 32] and references therein) and do not exceed 3–8% for the used crystal thickness. A specially designed helium cryostat [30] made it possible to measure the resistance after irradiation in the temperature range of 10 K < *T* < 300 K.

## 3. Results and discussion

### 3.1. Resistance and critical temperature

Figure 1 shows the temperature dependences of the resistivity $\rho(T) = \rho_{ab}(T)$ of the optimally doped (OD) single crystal YBa$_2$Cu$_3$O$_{7-\delta}$ with $T_c$ = 91.66 K ($\varphi_1$ = 0) and oxygen index 7 − δ ≈ 6.94, measured before irradiation ($\varphi_1$ = 0, gray dots) and for different irradiation doses $\varphi_2$ = 1.4·10$^{18}$ e/cm$^2$ (blue dots), $\varphi_3$ = 4.3·10$^{18}$ e/cm$^2$ (yellow dots), and $\varphi_4$ = = 8.8·10$^{18}$ e/cm$^2$ (turquoise dots). Thus, in fact, we are examining four different samples, the parameters of which are determined by the radiation dose. As usual, $T_c$ of our samples was determined by extrapolating the linear portion

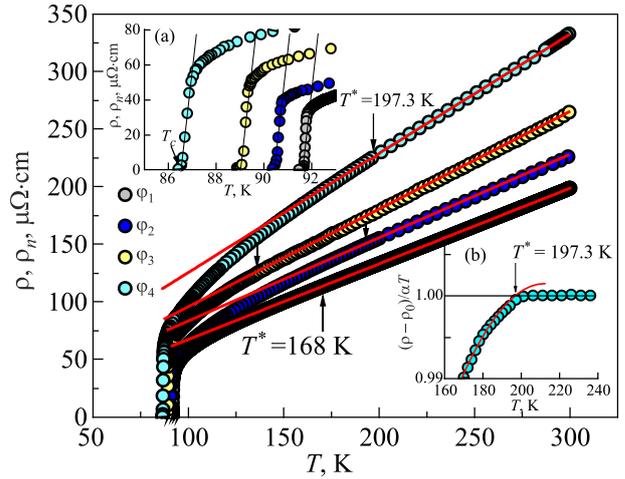

*Fig. 1.* (Color online) Temperature dependences of the resistivity of an optimally doped YBa$_2$Cu$_3$O$_{7-\delta}$ single crystal (7 − δ ~ 6.94) at different irradiation doses $\varphi_1$ = 0, $\varphi_2$ = 1.4·10$^{18}$ e/cm$^2$, $\varphi_3$ = = 4.3·10$^{18}$ e/cm$^2$, $\varphi_4$ = 8.8·10$^{18}$ e/cm$^2$. Red lines denote $\rho_N(T)$. Arrows designate temperatures $T^*$. Inset (a) is resistive transitions to the superconducting state at different $\varphi$, which determine $T_c$. Inset (b) is a more accurate determination of $T^*$ using the criterion $(\rho(T) - \rho_0)/\alpha T = 1$ [38] for $\varphi_4$ = 8.8·10$^{18}$ e/cm$^2$. The red curve is the polynomial fit data below $T^*$.

$\rho(T)$ at a SC transition to its intersection with the *T* axis (see inset (a) in Fig. 1). The $\Delta T_c$ = 0.25 K ($\varphi_1$ = 0) and $\Delta T_c$ = 0.5 K ($\varphi_4$ = 8.8·10$^{18}$ e/cm$^2$) determined in this way are very narrow, which indicates the good quality of our single crystals and the absence of additional phases with different $T_c$ in the samples. This is confirmed by the fact that the residual resistance ratio, $RRR = R_{300\,K}/R_{100\,K}$ = 3.5, and the residual resistivity, $\rho_0$, of the unirradiated sample close to zero (refer to Table 1), as is observed in well-structured YBCO films [12]. And although δ$T_c$ increases almost twofold under irradiation, the resistive transition still remains rather abrupt, which indicates a uniform distribution of induced defects (disorder) over the sample volume [20]. On the inset (b) to Fig. 1, as an example, the refined definition of $T^*$ for $\varphi_4$ = 8.8·10$^{18}$ e/cm$^2$ is given using the criterion $(\rho(T) - \rho_0)/\alpha T = 1$, where $\alpha = d\rho/dT$ [38]. The red curve is the polynomial fit data below $T^*$. The point of its intersection with the straight line gives $T^*$. This approach makes it possible to determine $T^*$ with an accuracy of ± 0.3 K.

Table 1. Resistive parameters of YBa$_2$Cu$_3$O$_{7-\delta}$ single crystal at different irradiation doses

| $\varphi$, 10$^{18}$ e/cm$^2$ | $\rho(300\,K)$, µΩ·cm | $\rho(100\,K)$, µΩ·cm | $\rho_0$, µΩ·cm | $d\rho/dT$, µΩ·cm/K | $T_c$, K | $T_c^{mf}$, K | $T_G$, K | $T_{01}$, K | $\Delta T_{fl}$, K |
|---|---|---|---|---|---|---|---|---|---|
| 0 | 200.0 | 57.6 | 1.9 | 0.66 | 91.7 | 91.73 | 91.8 | 94.2 | 2.4 |
| 1.4 | 226.0 | 63.6 | 11.2 | 0.72 | 90.5 | 90.61 | 90.6 | 93.6 | 3.0 |
| 4.3 | 265.0 | 83.4 | 10.3 | 0.84 | 89.0 | 89.21 | 89.3 | 92.8 | 3.5 |
| 8.8 | 332.0 | 100.6 | 21 | 1.03 | 86.6 | 86.67 | 86.7 | 96.8 | 10.1 |

Table 2. Parameters of fluctuation conductivity of YBa$_2$Cu$_3$O$_{7-\delta}$ single crystal at different irradiation doses

| $\varphi$, $10^{18}$ e/см$^2$ | $d_{01}$, Å | $\Delta\ln\sigma'$ | $\xi_c(0)$, Å | $\xi_{ab}(0)$, Å | $\xi_{ab}(T_{pair})$, Å |
|---|---|---|---|---|---|
| 0 | 2.2 | 0.68 | 0.37 | 10.0 | 17.7 |
| 1.4 | 2.9 | 0.89 | 0.50 | 13.5 | 20.4 |
| 4.3 | 3.2 | 0.40 | 0.70 | 18.9 | 31.6 |
| 8.8 | 3.6 | 0.21 | 1.20 | 29.6 | 39.2 |

As can be seen from the figure, at $T > T^*$ $\rho(T)$ of all samples is a linear function of temperature with a slope $d\rho/dT = 0.66$ μΩ·cm/K ($\varphi_1 = 0$) and $d\rho/dT = 1.03$ μΩ·cm/K ($\varphi_4 = 8.8 \cdot 10^{18}$ e/cm$^2$), respectively. The slope was calculated by the method of computer approximation of the experimental curves above $T^*$ and confirmed the linear behavior of $\rho(T)$ with a root-mean-square error of $0.023 \pm 0.002$ over a given temperature range for all samples. Unlike Ref. 20, the slope of the experimental curves is gradually increases under irradiation (Table 1). In other words, the scattering rate of normal charge carriers gradually increases with increasing $\varphi$. This is the experimental fact that we are trying to discuss and explain. In [20], the slope $d\rho/dT$ did not change with the irradiation dose, indicating the fulfillment of the Matthiessen rule, which in our case is not fully satisfied. The issue of deviation of experimental data from the Matthiessen rule (MR) has long been widely discussed in the literature [39]. General conclusion: deviation from MR is the rule, strict adherence to the MR is an exception (zero approximation). We believe that the discrepancy with the results of [20, 40–42] is most likely due to the structural features of each single crystal under study, which interacts differently with radiation defects. In our case, it is believed that it is TBs that are responsible for the observed shape of the resistivity curves under irradiation. Notably, when Giapintzakis *et al.* studied the effect of irradiation with electrons with an energy of 0.35 MeV on YBCO single crystals without TBs, then, as expected, the Matthiessen rule was observed [42]. Thus, the degree of change in crystal parameters depends on both the impurity composition of the initial samples and the energy of the incident particles [18–20, 29–33].

According to estimates in previous studies [18–20, 32] (and references therein), 1 MeV electrons cause displacements in YBa$_2$Cu$_3$O$_{7-\delta}$ of any of the four types of atoms O, Cu, Y, and Ba. Within the simple relativistic kinematic relation for electron-atom scattering it was found that the minimum electron energies required for displacement of O, Cu, Y, and Ba are 129, 413, 532, and 730 keV, respectively, assuming a threshold of 20 eV recoil energy [32]. Thus our electron irradiation with energy 2.5 MeV does produce a lot of defects in YBCO due to displacement of all atoms. Depending on the defect's location within the unit cell as well as the size and concentration of the point or small cluster defects, the properties of the material can be altered. In addition, the observed variation in the dose dependence of $\rho$, $d\rho/dT$, $J_c$ /$J_{c0}$ etc. from crystal to crystal [29–33] can be explained by the fact that each crystal has a different preirradiation defect state. The *radiation-induced point defects react with the preirradiation defects and change them*. In our case, such preirradiation defects are the TBs. As mentioned above, the density of charge carriers at the TBs is reduced, whereas it is somewhat increased just near TBs, compare with the other sample volume [17, 21]. Thus, TBs create an inhomogeneous charge distribution over the volume of the unirradiated sample. According to [32], TBs should be affected by radiation defects. As a consequence, the presence of radiation defects with electrical activity in the crystal leads to a decrease in the concentration of charge carriers and mobility, an increase in the scattering rate, especially in the normal state, and, as a consequence, to a decrease in $T_c$ and an increase in $\rho$ and $d\rho/dT$ observed in experiment. Moreover, in our opinion, this should also lead to the elimination of the effect of twins and TBs and equalization of the charge distribution over the sample.

In our experiment, electron irradiation leads to a noticeable increase in $\rho$ and the coherence length along the $c$ axis, $\xi_c$ (0), as well as to the expected decrease in $T_c$ (refer to Tables 1 and 2). In addition, a corresponding increase in the PG opening temperature $T^*$ is observed (refer to Table 3), which is consistent with the phase diagram of cuprates [8, 9, 17, 40, 43], namely the lower $T_c$ the greater $T^*$.

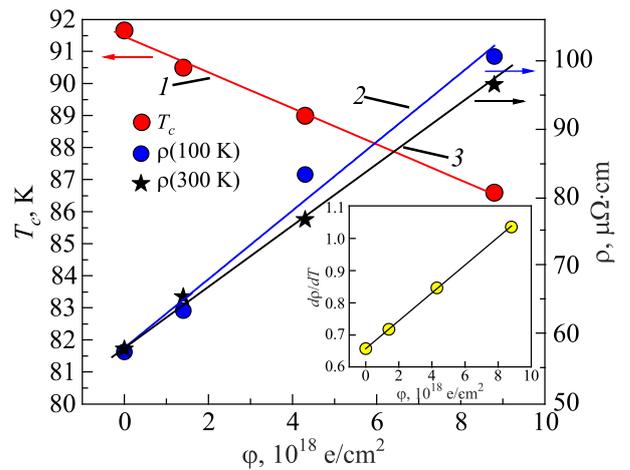

*Fig. 2.* (Color online) Dependences of $T_c$ (*1*), $\rho$(100 K) (*2*), and $\rho$(300 K) (*3*) on the irradiation dose $\varphi$ for the YBa$_2$Cu$_3$O$_{6.94}$ single crystal. The data for $\rho$(300 K) are multiplied by 0.29. Inset shows the linear dose dependence of $d\rho/dT$. Solid lines are guidance for eye.

Table 3. Parameters of the pseudogap analysis of $YBa_2Cu_3O_{7-\delta}$ single crystal at different irradiation doses

| $\varphi$, $10^{18}$ e/cm² | $T^*$, K | $T_{pair}$, K | $\varepsilon_{c0}^*$ | $A_4$ | $D^*$ | $\Delta^*(T_G)$, K | $\Delta^*(T_{pair})$, K |
|---|---|---|---|---|---|---|---|
| 0 | 168.0 | 121 | 0.16 | 6.1 | 5 | 229.0 | 207.6 |
| 1.4 | 193.0 | 130 | 0.23 | 8 | 5 | 225.4 | 180.0 |
| 4.3 | 136.4 | 121 | 0.16 | 13 | 5 | 216.4 | 189.7 |
| 8.8 | 197.3 | 136 | 0.23 | 20 | 4.8 | 202.7 | 236.0 |
| F1 [61] | 203.0 | 133 | 0.23 | 20 | 5 | 217.0 | 235.0 |

The dependences of $T_c$, $d\rho/dT$, $\rho$ (300 K) and $\rho$ (100 K) on the irradiation dose $\varphi$ are shown in Fig. 2. As in [18, 20], $T_c$ decreases linearly with increasing $\varphi$ (Fig. 2, line *1*). The observed $\Delta T_c \sim 5$ K is close to the data of [18] ($\Delta T_c \sim 3$ K), and is determined both by the features of the single crystal structure and by the irradiation intensity. Accordingly, both $d\rho/dT$ (inset in Fig. 2) and $\rho$(300 K) (straight line *2*) increase linearly with $\varphi$ (Table 1), despite the fact that at $\varphi_3 = 4.3 \cdot 10^{18}$ e/cm² a sharp unexpected decrease in $T^*$ from 193 to 136.4 K is observed (Fig. 1 and Table 3). However, the dependence of $\rho$ (100 K) on $\varphi$ (blue dots) deviates somewhat from the straight line just at $\varphi_3 = 4.3 \cdot 10^{18}$ e/cm² also suggesting the peculiar behavior of the crystal at this $\varphi$. As expected, the dose dependences of both FLC and especially PG show more pronounced features at $\varphi_3 = = 4.3 \cdot 10^{18}$ e/cm² and allow to get much more information, as will be discussed in detail below.

### 3.2. Fluctuation conductivity

Within the Nearly Antiferromagnetic Fermi–Liquid (NAFL) model [43] it was proven that a linear $\rho(T)$ above $T^*$ is an integral feature of the normal state of cuprates (e.g., YBCO), which is characterized by the stability of the Fermi surface. At $T \leq T^*$ $\rho(T)$ deviates downward from the linearity, resulting in appearance of the excess conductivity $\sigma'(T)$:

$$\sigma'(T) = \sigma(T) - \sigma_N(T) = \frac{1}{\rho(T)} - \frac{1}{\rho_N(T)}, \quad (1)$$

where $\rho_N(T) = \alpha T + \rho_0$ is the resistivity of the sample in the normal state, extrapolated to the region of low temperatures. As before, $\alpha = d\rho/dT$ determines the slope of the linear dependence $\rho_N(T)$, and $\rho_0$ is the residual resistance cut off by this line on the $Y$ axis at $T = 0$ [12, 44–46]. It is worth to note that at $T = T^*$ not only $\rho(T)$ deviates downward from linearity but also the density of states (DOS) at the Fermi level begins to gradually decrease, which means the opening of PG [11, 47, 48]. In addition, at $T = T^*$ the Fermi surface is believed to change [9, 10], most likely because of the formation of the local pairs (LPs) just below $T^*$ [12, 27, 28]. Thus the proper determination of $T^*$ is of primary importance for the FLC and PG analysis. Fortunately, the above method for finding $T^*$ [Fig. 1, inset (b)] allows a quite well way of determining $T^*$ with sufficient accuracy.

In accordance with modern concepts [5, 27, 28, 41–46, 49–51], the small coherence length in combination with the quasi-layered structure of HTSCs leads to the formation of a noticeable, in comparison with conventional superconductors, range of SC fluctuations, $\Delta T_{fl}$, above $T_c$. In this range, fluctuating Cooper pairs (CPs) behave in a good many ways like ordinary SC pairs, but without long-range order (the so-called "short-range phase correlations") [1–5, 27, 28, 49], and the excess conductivity $\sigma'(T)$ obeys the classical fluctuation theories [52–56]. Thus, the fluctuation conductivity is only a part of the total $\sigma'(T)$ and is realized in a relatively narrow interval of SC fluctuations near $T_c$. This interval is $\Delta T_{fl} = T_{01} - T_G \leq 20$ K in all HTSCs including even FeSe [57, 58], $T_G$ is the Ginsburg temperature, down to which the Bogolyubov's mean-field theory works. The diversity of $\Delta T_{fl}$ is determined by the change in oxygen stoichiometry, the presence of impurities and/or structural defects, which should also have a significant effect on $\sigma'(T)$ and, accordingly, on the implementation of various models for describing the FLC above $T_c$ [12, 44, 45, 51–54]. However, it should be emphasized that the effect of high-energy electron irradiation on $\sigma'(T)$ has not yet been studied.

To estimate the FLC within the framework of the LPs model [5, 12, 27, 28], it is first necessary to determine the critical temperature in the mean-field approximation, $T_c^{mf}$ [12, 21, 25], which limits the range of critical fluctuations around $T_c$, in which the SC order parameter $\Delta < k_B T$ [59, 60]. The $T_c^{mf}$ is an important parameter of both FLC and PG analysis, since it determines the reduced temperature $\varepsilon = (T - T_c^{mf})/T_c^{mf}$, which is included in all equations. In HTSCs near $T_c$, the FLC is always described by the Aslamazov–Larkin (AL) [55] equation for any 3D system [12, 25, 45]:

$$\sigma'_{AL3D} = C_{3D} \frac{e^2}{32\hbar\xi_c(0)} \varepsilon^{-1/2}. \quad (2)$$

To determine $T_c^{mf}$ we use an approach proposed by Beasley *et al.*: from Eq. 2, $\sigma'^{-2}(T) \sim \varepsilon \sim T - T_c^{mf}$ and is zero when $T = T_c^{mf}$ [60]. The result is shown in Fig. 3, using $\varphi_1 = 0$ as an example. In addition to $T_c^{mf} = 91.73$ K, $T_c$, the Ginzburg temperature $T_G > T_c^{mf}$, and the 3D–2D crossover temperature $T_0$ are also shown. Using the same approach, $T_c^{mf}$ were obtained for all $\varphi$ (refer to Table 1).

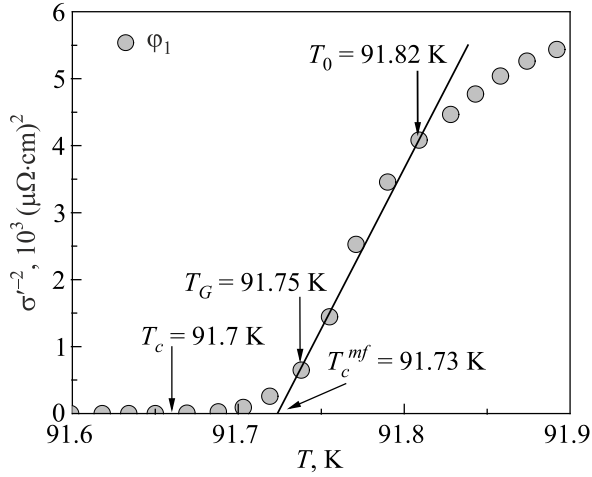

*Fig. 3.* Temperature dependence of the inverse square of the excess conductivity, $\sigma'^{-2}(T)$, for the OD single crystal of YBa₂Cu₃O₆.₉₄, which determines $T_c^{mf} = 91.73$ K at $\varphi_1 = 0$. The arrows also indicate $T_c$, the Ginzburg temperature $T_G$, and the 3D–2D crossover temperature $T_0$.

For further analysis of the FLC, it is necessary to plot the dependences of ln σ′ on ln ε for each radiation dose and compare the results with fluctuation theories [12, 44, 45] (refer to Fig. 4). For a better understanding of the results, it should be taken into account that the unusual properties of cuprates are largely determined by the very short coherence length, which is largely anisotropic. The specific temperature behavior of FLC in cuprates is determined by the coherence length along the c axis, $\xi_c(T) = \xi_c(0)\varepsilon^{-1/2}$ [12, 45, 56, 61], where $\xi_c(T^*) \sim 1$ Å, which is comparable to the unit cell size of YBCO $d = 11.68$ Å [62]. Near $T_c$, $\xi_c(T) > d$, and fluctuating CPs interact over the entire volume of the sample, forming a three-dimensional (3D) state [12, 44, 51, 54, 61, 63], which is described by the AL theory [Eq. (2)]. These are the red dashed lines with a slope λ = −1/2, marked as 3D–AL in the figure. The corresponding parameters of all samples at various φ are listed in Tables 1–3.

Since $\xi_c(T)$ decreases with increasing $T$, then at $T > T_0$, which limits the range of 3D fluctuations from above, $\xi_c(T) < d$, and 3D state is lost. However, as before, $\xi_c(T) > d_{01} \sim 3.4$ Å, which is the distance between the internal conducting CuO₂ planes [62] and thus, connects the CuO₂ planes by the Josephson interaction. This is the range of 2D SC fluctuations which is well described by the 2D–MT fluctuation contribution (Fig. 4, blue curves) calculated within the Hikami–Larkin theory [56] with the parameters given in Table 2:

$$\sigma'_{MT} = C_{2D}\frac{e^2}{8d\hbar}\frac{1}{1-\alpha/\delta}\ln\left((\alpha/\delta)\frac{1+\alpha+\sqrt{1+2\alpha}}{1+\delta+\sqrt{1+2\delta}}\right)\varepsilon^{-1}, \quad (3)$$

where $\alpha = 2[\xi_c(0)/d]^2\varepsilon^{-1}$ is the coupling parameter,

$$\delta = \beta\frac{16}{\pi\hbar}\left[\frac{\xi_c(0)}{d}\right]^2 k_B T\tau_\varphi \quad (4)$$

is the pair-breaking parameter, $\tau_\varphi\beta T = \pi\hbar/8k_B\varepsilon_0 = A/\varepsilon_0$ is the lifetime of the fluctuating CPs and $A = 2.998\cdot10^{-12}$ sK. The factor β = 1.203($\ell/\xi_{ab}$), where $\ell$ is the mean free path, corresponds to the case of the clean limit ($\ell > \xi$), which is typical for HTSCs [12, 53, 54, 61]. On a double logarithmic scale, these are solid blue curves (2D–MT) in Fig. 4, which perfectly describe the data in the range from $T_0$, marked as ln ε₀ and $T_{01}$ (ln ε₀₁ in the figure), which limits the range of SC fluctuations from above. Therefore, it is $\xi_c(T)$ that governs the FLC behavior in HTSCs near $T_c$.

Thus, at $T = T_0$, the 3D–2D (AL–MT) crossover occurs [45, 54, 56, 63]. Obviously, at $T_0$ we have $\xi_c(T_0) = \xi_c(0)\varepsilon_0^{1/2} = d = 11.68$ Å, which allows the determination of the coherence length along the c axis

$$\xi_c(0) = d\sqrt{\varepsilon_0}. \quad (5)$$

After acquiring $T_0 \approx 91.8$ K (ln ε₀ ≈ −6.9), from the Eq. (5) we find $\xi_c(0) = (0.37 \pm 0.02)$ Å ($\varphi_1 = 0$), which is a typical value of $\xi_c(0)$ for optimally doped YBCO single crystals with twins and the same $T_c = 91.7$ K [64].

At $T_{01}$, corresponding to ln ε₀₁ at Fig. 4, the experimental data completely deviate from the theory. This is due to the fact that, above $T_{01}$, the fluctuating CPs are located inside the conducting $CuO_2$ planes and are no longer connected by a correlation interaction [12, 44, 54], since now $\xi_c(T) < d_{01}$. Thus, it is clear that $\xi_c(T_{01}) = d_{01}$. To find $d_{01}$, we use the experimental fact that $\xi_c(0)$ has already been determined from the crossover temperature $T_0$. Consequently, the condition

$$\xi_c(0) = d\sqrt{\varepsilon_0} = d_{01}\sqrt{\varepsilon_{01}} = (0.39 \pm 0.1) \text{ Å}$$

is met ($\varphi = 0$). Since $d = c = 11.68$ Å and $T_{01} \approx 94.2$ K ($\ln \varepsilon_{01} \approx -3.6$), we obtain: $d_{01} = d\sqrt{\varepsilon_0 / \varepsilon_{01}} = (2.2 \pm 0.1)$ Å for $\varphi = 0$. Somewhat surprisingly, this is noticeably smaller than the interplanar distance $d_{01} \approx 3.4$ Å in YBCO without TBs [62]. Thus, it can be concluded that TBs somehow affect the distance between the conducting $CuO_2$ planes in YBCO, denoted as $Cu(2)–Cu(2)$ in [62]. Performing a similar analysis for all other irradiation dose $\varphi$, we obtain the values of $\xi_c(0)$ and $d_{01}$ for all four samples (Table 2). For example, at $\varphi_4 = 8.8 \cdot 10^{18}$ e/cm$^2$, for which $T_0 \approx 87.5$ K ($\ln \varepsilon_0 \approx -4.73$) and $T_{01} \approx 93.9$ K ($\ln \varepsilon_{01} \approx -2.47$) (refer to Fig. 4), $\xi_c(0) = (1.2 \pm 0.1)$ Å and $d_{01} = (3.6 \pm 0.1)$ Å are obtained. Now $d_{01}$ is close to that found for YBCO without TBs [62], which indicates the correct determination of $\varepsilon_{01}$. This also means that $d_{01}$ increased by ~ 1.7 times, suggesting an improvement in the structure of the sample after irradiation. Accordingly, $\xi_c(0)$ increases by a factor of ~3 upon irradiation, which seems reasonable, since in the theory of superconductivity $\xi \sim 1/T_c$ [59].

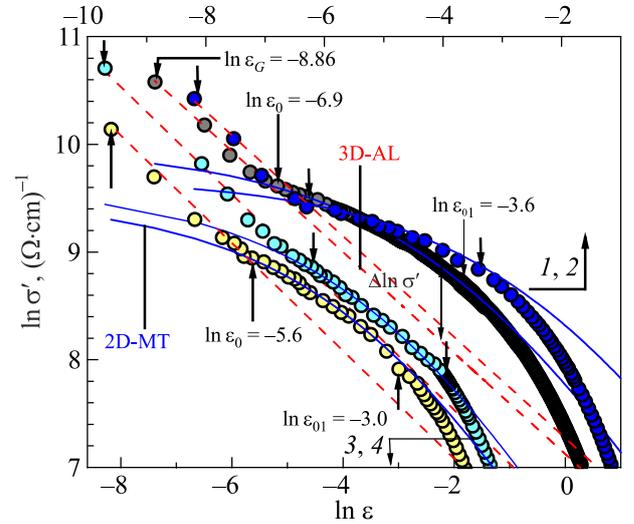

*Fig. 4.* (Color online) Dependences of ln σ′ on ln ε of the YBa₂Cu₃O₆.₉₄ single crystal at $\varphi_1 = 0$ (gray dots), $\varphi_2 = 1.4\cdot10^{18}$ e/cm² (blue dots), $\varphi_3 = 4.3\cdot10^{18}$ e/cm² (yellow dots), and $\varphi_4 = 8.8\cdot10^{18}$ e/cm² (turquoise dots) in comparison with fluctuation theories: 3D–AL (red dashed lines) and 2D–MT (solid blue curves). ln ε$_G$ defines the Ginzburg temperature $T_G$, ln (ε₀) defines the 3D−2D crossover temperature $T_0$, and ln (ε₀₁) defines $T_{01}$. The scales along the X axis are shifted for readability: at $\varphi_1 = 0$ and $\varphi_2 = 1.4\cdot10^{18}$ e/cm², the upper scale is used. The numbers near the arrows are marked for $\varphi_1 = 0$ and $\varphi_3 = 4.3\cdot10^{18}$ e/cm², except for ln ε$_G$ = − 8.17 for $\varphi_3$.

All characteristic temperatures are designated by arrows in the figure. The long arrows are labeled for $\varphi_1 = 0$ and $\varphi_3 = 4.3 \cdot 10^{18}$ e/cm$^2$, with the exception of the leftmost point for $\varphi_3$, which is $\ln \varepsilon_G = -8.17$ (see lower $X$ scale). The short arrows for $\varphi_2 = 1.4 \cdot 10^{18}$ e/cm$^2$ and $\varphi_4 = 8.8 \cdot 10^{18}$ e/cm$^2$ are not labeled for better readability. In general, Fig. 4 shows that at all $\varphi$ the agreement of the data with the AL and MT theories is very good. This fact suggests that the normal state of the samples and $T^*$'s are chosen correctly. This also suggests that up to $T_{01}$, the wave function phase stiffness of the order parameter has to be maintained and the superfluid density, $n_s$, is nonzero [27, 28, 65]. As a result, in this temperature range, the fluctuating CPs largely behave like the SC but non-coherent pairs, as mentioned above.

However, some specific features of the FLC behavior in YBCO single crystals under irradiation worth discussing in more detail. As already noted, at $\varphi_1 = 0$ the dependence of $\ln \sigma'$ on $\ln \varepsilon$ is quite usual with the behavior of 3D–AL near $T_c$ and 2D–MT above $T_0$ (Fig. 4). However, the deviation of 2D–MT fluctuations from 3D–AL (indicated by the double arrow as $\Delta \ln \sigma'$ in Fig. 4) is relatively large, namely, $\Delta \ln \sigma' = 0.68$. Such enhancement of 2D–MT fluctuations is characteristic of optimally doped single crystals containing a certain number of defects, mainly in the form of TBs [21, 64].

With a slight increase in dose up to $\varphi_2 = 1.4 \cdot 10^{18}$ e/cm$^2$, the absolute value of $\sigma'$ also slightly increases in both 3D and 2D states (blue dots in the figure). Accordingly, the range of SC fluctuations, $\Delta T_{\text{fl}}$, which is controlled by $T_{01}$ ($\ln \varepsilon_{01}$ in the figure), also increases (refer to Table 1). The results indicate a strong effect of radiation defects on the properties of the crystal, but the effect of TBs also remains strong. This is evidenced by a further increase in the deviation of fluctuations of 2D–MT over 3D–AL, with $\Delta \ln \sigma'$ taking the largest value of ~ 0.89 (Table 2). In addition, $d_{01} = 2.9$ Å is still small.

As the dose increases to $\varphi_3 = 4.3 \cdot 10^{18}$ e/cm$^2$, the absolute value of FLC noticeably decreases (yellow dots in Fig. 4). The FLC shift can be estimated by comparing the results at different $\varphi$ with the same $\ln \varepsilon$. For example, at $\varphi_1$, one can take the value of FLC at $T = T_0$, for which $\ln \varepsilon_0 = -6.9$, $\ln \sigma' \approx 9.66$, and $\sigma'(T_0) \approx 15650$ ($\Omega \cdot$cm)$^{-1}$, respectively. At $\varphi_3$, $\ln \varepsilon = -6.9$ corresponds to the third point from the left, for which $\ln \sigma' \approx 9.25$ and $\sigma' \approx 10400$ ($\Omega \cdot$cm)$^{-1}$. Ultimately, $\sigma'(\varphi_1) / \sigma'(\varphi_3) \approx 1.5$ times. The data obtained indicate that the displacement of atoms and the resulting defects strongly affect $\sigma'$. Accordingly, the deviation $\Delta \ln \sigma' = 0.40$, that is, it decreases by more than 2 times, while $\Delta T_{\text{fl}}$ (Table 1) and $d_{01}$ (Table 2) continue to grow. Interestingly, $d_{01}$ is almost the same as in YBCO without TBs [62].

With a further increase in the dose by more than two times to $\varphi_4 = 8.8 \cdot 10^{18}$ e/cm$^2$, the absolute value of the FLC unexpectedly almost recovers (Fig. 4, turquoise dots). At the same time, the deviation $\Delta \ln \sigma' = 0.21$ takes the smallest value, while $\Delta T_{\text{fl}}$ increases by more than 2 times (Table 1) and $d_{01}$ acquires the largest value ~ 3.6 Å (Table 2). Interestingly, all of the above parameters are characteristic of well-structured YBCO films. Thus, quite unexpectedly the obtained dependence of $\ln \sigma'$ on $\ln \varepsilon$ takes the shape which is typical for a well-structured YBCO film with a close $T_c \approx 87.4$ K [45, 61] indicated in Table 3 as F1. Moreover, in this case, as for the F1 film, the scale factor $C_{3D} = 1$ in Eq. (2), which indicates full agreement between the data and the AL theory. Thus, when irradiated with high-energy electrons with the highest dose $\varphi_4$, our single crystal behaves almost like *a defect-free sample of YBCO*.

To account for this very surprising result, we must remember that our original sample contains a number of twins with pronounced TBs. The TBs create an inhomogeneous charge distribution over the volume of the unirradiated sample. As discussed above, incident 2.5 MeV electrons effectively displace all atoms in YBCO, deeply affecting TBs. Finally, taking into account the results obtained, we can conclude that *the irradiation defects completely equalized the distribution of charge, that is, virtually eliminated the effect of twins and twin boundaries in the sample*. In fact, the *greater the disorder caused by defects, the more isotropic the sample* [19, 66, 67]. Naturally, we expected to find confirmation of this result by analyzing the temperature dependences of the pseudogap at different $\varphi$.

### 3.3. Temperature dependence of the pseudogap

If there was no PG in the HTSCs, the temperature dependence of $\rho(T)$ would remain linear down to $T_c$ [5, 6, 12, 44, 51]. This means that the excess conductivity $\sigma'(T)$ must contain information on the magnitude and temperature dependence of the PG. Obviously, to obtain such information, it is necessary to have an equation that would describe $\sigma'(T)$ over the entire temperature range from $T^*$ to $T_G$ and contain the explicit parameter PG, $\Delta^*(T)$. However, up to now any rigorous theory of HTSCs to describe $\sigma'(\varepsilon)$ above $T_{01}$ is still missing. Therefore, the analysis was carried out using equation proposed in [61]

$$\sigma'(T) = A_4 \frac{e^2 \left(1 - \dfrac{T}{T^*}\right) \exp\left(-\dfrac{\Delta^*}{T}\right)}{16\hbar\xi_c(0)\sqrt{2\varepsilon_{c0}^*\ \sinh\left(2\dfrac{\varepsilon}{\varepsilon_{c0}^*}\right)}}, \qquad (6)$$

where $(1 - T/T^*)$ determines the number of pairs arising at $T \leq T^*$, and $\exp(-\Delta^*/T)$ gives the number of pairs destroyed by thermal fluctuations below $T_{\text{pair}}$ [12, 61]. The Eq. (6) is based on ideas from Leridon *et al.* [68] and Prokof'ev *et al.* [69] but markedly modified to provide the best fit for $\sigma'(T)$ over the entire temperature range from $T^*$

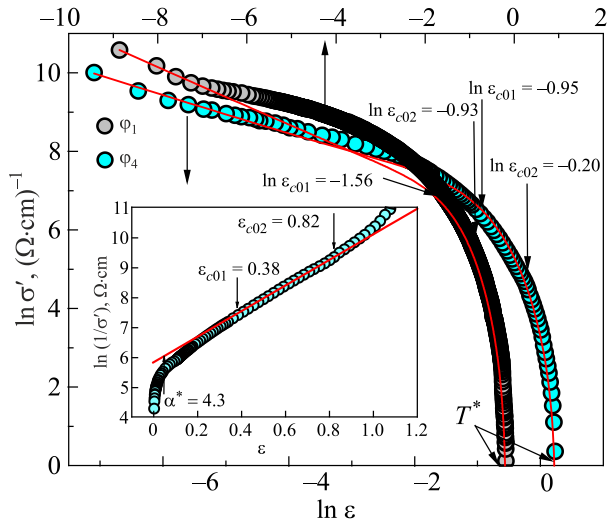

*Fig. 5.* (Color online) Dependences of $\ln\sigma'$ vs $\ln\varepsilon$ (dots) of the YBa$_2$Cu$_3$O$_{6.94}$ single crystal in the entire temperature range from $T^*$ down to $T_G$ at $\varphi_1 = 0$ (gray dots) and $\varphi_4 = 8.8\cdot10^{18}$ e/cm$^2$ (turquoise dots). Red curves are approximation of experimental data by Eq. (6) with a set of parameters given in the text. Inset: $\ln(1/\sigma')$ as a function of $\varepsilon$ at $\varphi_4 = 8.8\cdot10^{18}$ e/cm$^2$. The straight red line denotes the linear part of the curve between $\varepsilon_{c01} = 0.38$ and $\varepsilon_{c02} = 0.82$. The corresponding values of $\ln(\varepsilon_{c01})$ and $\ln(\varepsilon_{c02})$ are indicated by arrows on the main panel for both $\varphi$ values. The slope $\alpha^* = 4.3$ determines the parameter $\varepsilon_{c0}^* = 1/\alpha^* = 0.23$ at $\varphi_4 = 8.8\cdot10^{18}$ e/cm$^2$. The $X$ scales are shifted for readability.

down to $T_G$. Solving Eq. (6) with respect to $\Delta^*(T)$, we obtain the equation for the PG

$$\Delta^*(T) = T \ln\Bigg[ A_4 \left(1 - \frac{T}{T^*}\right) \frac{1}{\sigma'(\varepsilon)} \frac{e^2}{16\hbar\xi_c(0)}$$

$$\times \frac{1}{\sqrt{2\varepsilon_{c0}^*\ \sinh\left(2\varepsilon/\varepsilon_{c0}^*\right)}} \Bigg], \qquad (7)$$

where $\sigma(\varepsilon)$ is the experimentally determined excess conductivity.

In addition to $T^*$, $T_c^{mf}$, $\xi_c(0)$, and $\varepsilon$ already defined above, both equations contain the theoretical parameter $\varepsilon_{c0}^*$ [68], the coefficient $A_4$, which has the same meaning as the *C*-factor in the theory of FLC, and $\Delta^*(T)$. Note that in the model of LPs, all parameters included in Eqs. (6) and (7) can be determined from experiments [12, 25, 61, 63, 70], as discussed below. According to Ref. 68, in the region $\ln\varepsilon_{01} < \ln\varepsilon < \ln\varepsilon_{02}$ (refer to Fig. 5) $\sigma'^{-1} \sim \exp(\varepsilon)$. Most likely, this behavior of excess conductivity is an intrinsic property of HTSCs [12, 61, 63, 68, 70]. As a result, in the temperature range $\varepsilon_{01} < \varepsilon < \varepsilon_{02}$ (124 K $< T <$ 158 K), $\ln\sigma'^{-1}$ is a linear function of $\varepsilon$ with a slope $\alpha^* = 4.3$, which defines the parameter $\varepsilon_{c0}^* = 1/\alpha^* = 0.23$ at $\varphi_4 = 8.8\cdot10^{18}$ e/cm$^2$, which is taken as an example (refer to the inset of Fig. 5). The same graphs were obtained for all irradiation doses, and reliable values of $\varepsilon_{c0}^*$ were found (refer to Table 3), which significantly affect the shape of the theoretical curves shown in Figs. 5 and 6, especially at high temperatures [25, 63, 70].

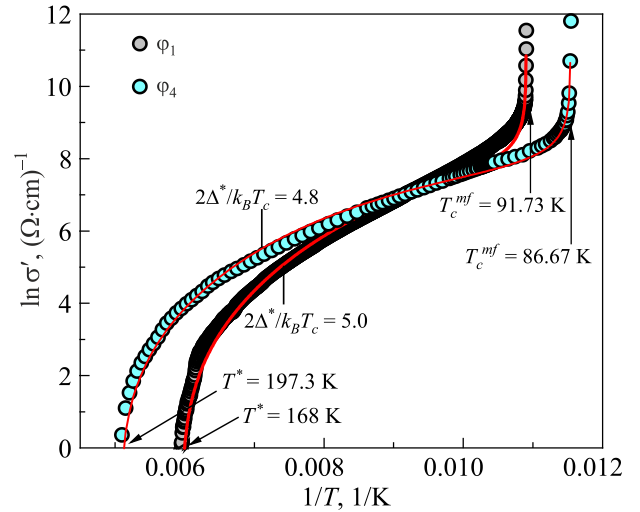

*Fig. 6.* (Color online) $\ln\sigma'$ as a function of $1/T$ of the YBa$_2$Cu$_3$O$_{6.94}$ single crystal in the entire temperature range from $T^*$ down to $T_c^{mf}$ at $\varphi_1 = 0$ (gray dots) and $\varphi_4 = 8.8\cdot10^{18}$ e/cm$^2$ (turquoise dots). Red curves are approximation of experimental data by Eq. (6) with a set of parameters given in the text. The best approximation is achieved at the value of the BCS ratio $D^* = 2\Delta^*(T_G)/k_BT_c = 5$ ($\varphi_1 = 0$) and $D^* = 4.8$ for $\varphi_4 = 8.8\cdot10^{18}$ e/cm$^2$.

To find $\Delta^*(T_G)$, the experimental values of excess conductivity are plotted in coordinates $\ln \sigma'$ vs $1/T$ [61, 69] (Fig. 6) and are well approximated by the theoretical dependences $\ln \sigma'(1/T)$ calculated by Eq. (6) (red curves in Fig. 6). With such a construction, the shape of the theoretical curves turns out to be very sensitive to the value of $\Delta^*(T_G)$ [25, 61, 63, 70]. It should be noted that in cuprates $\Delta^*(T_G) = \Delta(0)$, which is a SC energy gap at $T = 0$ [61, 71, 72]. The fact allows to determine the Bardeen–Cooper–Schriefer (BCS) ratio $D^* = 2\Delta(0)/k_B T_c = 2\Delta^*(T_G)/k_B T_c$, and hence the value $\Delta^*(T_G)$ at all irradiation doses. The best approximation is achieved at the value of $D^* = 2\Delta^*(T_G)/k_B T_c = 5$ for doses from $\varphi_1 = 0$ and up to $\varphi_3 = 4.3 \cdot 10^{18}$ e/cm², and $D^* = 4.8$ for $\varphi_4 = 8.8 \cdot 10^{18}$ e/cm² (Table 3). $D^* = (5 \pm 0.2)$ is a typical value for YBCO, suggesting the strong coupling limit for HTSCs [73]. The curves for $\varphi_2 = 1.4 \cdot 10^{18}$ e/cm² and $4.3 \cdot 10^{18}$ e/cm² are not shown so as not to clutter the graph.

Now the coefficient $A_4$ can be determined. For this, using Eq. (6) with $\Delta^* = \Delta^*(T_G)$, the dependence $\sigma'(T)$ is calculated with the parameters already found and, selecting $A_4$, is combined with the experiment in the region of 3D–AL fluctuations, where $\ln \sigma'$ is a linear function of $\ln \varepsilon$ with the slope $\lambda = -1/2$ [12, 54–56, 61, 63] (Fig. 5, red curves). Notably, in the range from $T_0$ ($\ln \varepsilon_0 = -6.9$) to $T_{c01}$ ($\ln \varepsilon_{c01} = -1.56$), the theoretical red curve for $\varphi_1 = 0$ deviates downward from the experimental data, which is typical for the optimally doped YBCO single crystals containing twins [64] or magnetic impurities in the sample [63]. Thus, at $\varphi_1 = 0$, the fit in the region of 2D fluctuations is not completely good, since our model does not take into account the influence of defects. However, the approximation of the experimental data by Eq. (6) at $\varphi_4 = 8.8 \cdot 10^{18}$ e/cm² (Fig. 5, corresponding red curve) is almost ideal. In fact, this confirms the above conclusion that a large number of defects cancels out the influence of the TBs and leads to isotropization of the sample [66, 67]. The $A_4$ values found for all irradiation doses are given in Table 3.

Having determined all the necessary parameters (refer to Tables 1, 2, and 3) we succeeded to plot the temperature dependences PG, $\Delta^*(T)$ for all radiation doses. It is worth to notice that in the PG analysis it is the $ab$ plane coherence length $\xi_{ab}(T) = \xi_{ab}(0)\varepsilon^{-1/2}$ that determines the LPs size and governs $\Delta^*(T)$ [12, 17]. Recall, that in cuprates the $\xi_{ab}(0) = \xi_{ab}(T^*) \sim 10$ Å, that is very short. As a result, at $T \leq T^*$ the LPs are in form of so-called strongly bound bosons (SBB), which obey the Bose–Einstein condensation (BEC) theory [28]. The SBB are very small and extremely tightly bound pairs that do not interact with each other and cannot be destroyed by any external influence. Along with decreasing temperature, $\xi_{ab}(T)$ and, consequently, the pair size increase, and at $T_{pair}$ LPs begin to interact with each other. With further decrease of $T$, the LPs begin to overlap and in the range of SC fluctuations behave like Cooper pairs which obey to BCS theory. Thus, in cuprates, there is a BEC–BCS crossover at $T_{pair}$ [28] first discovered experimentally in [61].

The result of our PG analysis is shown in Fig. 7. For example, the curve $\Delta^*(T)$ for $\varphi_1 = 0$ is calculated with the following set of parameters: $T^* = 168$ K, $T_c^{mf} = 91.73$ K, $\xi_c(0) = 0.37$ Å, $\varepsilon_{c0}^* = 0.16$, and $A_4 = 6.1$ and is represented by gray dots in Fig. 7. The jumps observed on all curves at high temperatures have no physical meaning but arise as a result of a change in the cooling rate of the samples when changing the rate of data recording. As expected from the FLC analysis, a rather unusual evolution of the shape of $\Delta^*(T)$ with increasing radiation dose was observed (Fig. 7).

At $\varphi_1 = 0$, $\Delta^*(T)$ has the shape characteristic of optimally doped OD YBCO single crystals with a relatively narrow maximum at $T_{pair} \approx 120$ K and a pronounced minimum at $T \approx T_{01} = 94.2$ K, below which $\Delta^*(T)$ is greatly increased. Interestingly, such a sharp increase in PG from $\Delta^*(T_{01}) \approx 161$ K up to $\Delta^*(T_G) = 229$ K (Table 3) in a very narrow interval of SC fluctuations $\Delta T_{fl} \sim 2.4$ K (Table 1) is characteristic feature of the PG behavior in all OD YBCO single crystals containing TBs [64].

At $\varphi_2 = 1.4 \cdot 10^{18}$ e/cm² (curve 2), the effect of radiation defects becomes already very noticeable. Accordingly, $\Delta^*(T_{pair})$ decreases markedly, showing a minimum value of 180K (refer to Table 3), most likely due to the observed increase in FLC (see Fig. 4). The entire $\Delta^*(T)$ curve shifts noticeably to the region of higher temperatures with $T_{pair} = 130$ K and $T^* = 193$ K. However, as follows from the phase diagram for YBCO [17, 40], the observed increase in $T^*$ up to 193 K at $T_c = 90.3$ K is too large and most likely can also be explained by the effect of irradiation. Nevertheless, as before, below $T_{01}$, a very sharp, almost 90 K, jump in $\Delta^*(T)$ up to $T_G$ is observed, and in a very narrow temperature range, which indicates that the influence of TBs is also still large. Interestingly, a similar evolution of the $\Delta^*(T)$ shape is observed when a YBCO single crystal with TBs is doped with ~ 20 % of Pr [64], which also produces the additional defects.

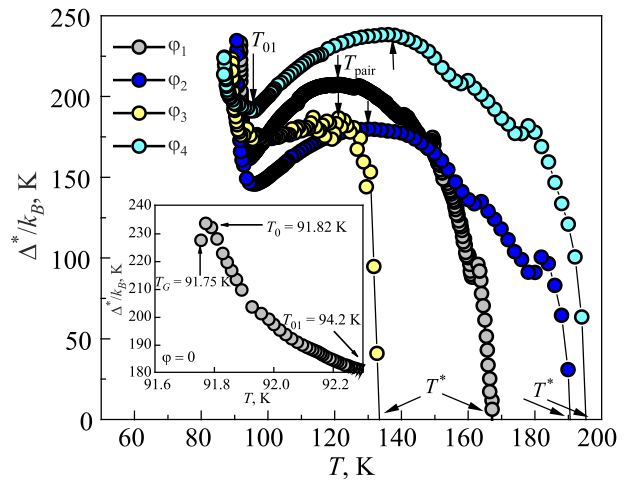

Fig. 7. (Color online) Temperature dependences of the pseudogap $\Delta^*(T)$ for the OD single crystal of $YBa_2Cu_3O_{6.94}$ at different irradiation doses: $\varphi_1 = 0$ (gray dots), $\varphi_2 = 1.4 \cdot 10^{18}$ e/cm² (blue dots), $\varphi_3 = 4.3 \cdot 10^{18}$ e/cm² (yellow dots), and $\varphi_4 = 8.8 \cdot 10^{18}$ e/cm² (turquoise dots), calculated using Eq. (7) with the parameters given in the text. Inset shows dependence $\Delta^*(T)$ at $\varphi_1 = 0$ in the range from $T_G$ to $T_{01}$. The arrows indicate the corresponding characteristic temperatures. Solid curves are guidance for eye.

However, unlike [64], when $\varphi$ increases ~ 3 times to $\varphi_3 = 4.3\cdot10^{18}$ e/cm$^2$, the $\Delta^*(T)$ is unexpectedly strongly compressed towards low $T$ (Figs. 7 and 8, yellow dots). As a result, all characteristic temperatures of the sample decreased sharply (refer to Table 3 and Fig. 9, curves *1* and *2*). Thus, we can conclude that due to the displacement of atoms by incident electrons, a specific system of defects arises in the crystal, which serve as effective pair-breaking centers [32]. Indeed, the observed sharp decrease in $T^*$ from 193 to 139 K is most likely due to the fact that such defects violate the phase coherence between electrons and prevent the formation of local pairs at $T \leq T^*$ [74]. As a result, $\Delta^*(T)$ acquires a rather specific shape, characteristic of OD YBCO single crystals with very low $T^*$, in which TBs are largely suppressed by annealing [70]. To verify this, we plotted in one figure $\Delta^*(T)$ for $\varphi_3 = 4.3\cdot10^{18}$ e/cm$^2$ (yellow dots), as well as $\Delta^*(T)$ for $P = 0$ GPa (red dots) and $P = 1$ GPa (blue squares) from [70] (refer to Fig. 8). As can be seen from the figure, the shape of all curves is completely the same. Below $T_{pair} = 121$ K, $\Delta^*(T)$ for $\varphi_3 = = 4.3\cdot10^{18}$ e/cm$^2$ becomes linear with the same positive slope $\alpha_1 \approx 0.53$, as is observed in [70]. Moreover, in this case curve $\varphi_3$ is approximately in the middle between $\Delta^*(T)$ obtained at $P = 0$ (red dots) and $P \approx 1$ GPa (blue squares).

Thus, we can conclude that *the induced defects most likely create an internal pressure $P \approx 0.6$ GPa in the YBCO crystal*, due to the displacement of individual atoms under irradiation, as noted above [18, 29, 31–33].

It was natural to expect that with a further increase in the irradiation dose, a similar form of the $\Delta^*(T)$ dependence will be obtained, with parameters even more suppressed by defects. However, surprisingly, this did not happen, but on the contrary: with an increase in $\varphi$ by a factor of ~2 to $\varphi_4 = 8.8\cdot10^{18}$ e/cm$^2$, $\Delta^*(T)$ unexpectedly takes the shape *typical of well-structured YBCO films* [61] *and untwined single crystals* [75]. Accordingly, $T^*$ increases up to 197.3 K (Fig. 9, curve *1*), while $T_{pair}$ (Fig. 9, curve *2*) and $\Delta^*(T_{pair})$ (Table 3) increase to 136 and 236 K, respectively. Moreover, all the found parameters of our sample at $\varphi_4 = 8.8\cdot10^{18}$ e/cm$^2$ turned out to be practically the same as for a well-structured YBCO film F1 with close $T_c = 87.4$ K, as shown in the last row of Table 3. Thus, as it follows from the literature (see [19, 66, 67] and references therein), a large number of radiation defects in the sample should *lead to "isotropization" of both the phonon spectrum and the charge distribution in the crystal*, which, rather, in total, is responsible for the observed shape of PG at $\varphi_4 = 8.8\cdot10^{18}$ e/cm$^2$. In other words, a large number of radiation defects in the sample reduces the effect of TBs, which create a nonuniform charge distribution in an unirradiated crystal. The found nontrivial dependence of PG on $\varphi$ is in full agreement with the dose dependence of FLC (Fig. 4). This allows us to repeat the above conclusion that *a large number of radiation defects completely equalized the charge distribution, that is, virtually eliminated the effect of all twins and TBs on the local pairs in the sample.*" In fact, the *greater the disorder caused by defects, the more isotropic the sample* [66, 67].

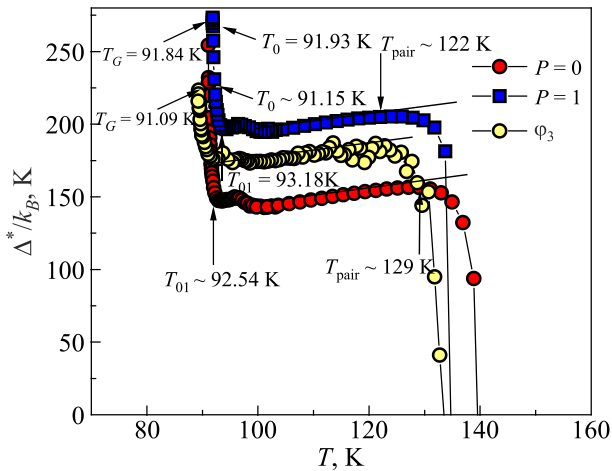

*Fig. 8.* (Color online) Temperature dependences of the pseudogap $\Delta^*(T)$ calculated in the LP model for the YBa$_2$Cu$_3$O$_{6.94}$ single crystal, in which TBs are largely suppressed by annealing, at $P = 0$ (red dots) and $P = 1$ GPa (blue squares) [70] compared with the result of current work at $\varphi_3 = 4.3\cdot10^{18}$ e/cm$^2$ (yellow dots). Arrows indicate the characteristic temperatures $T_{pair}$, $T_{01}$, $T_0$, and $T_G$ for the samples from Ref. 70. The corresponding temperatures for the sample at $\varphi_3 = 4.3\cdot10^{18}$ e/cm$^2$ are given in Tables 1 and 3. Solid curves are guidance for eye.

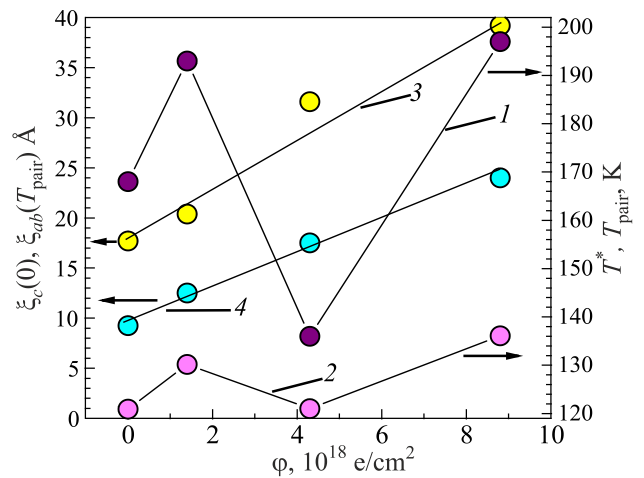

*Fig. 9.* (Color online) Dose dependences of the temperatures $T^*$ (curve *1*) and $T_{pair}$ (curve *2*), as well as the coherence lengths $\xi_{ab}(T_{pair})$ (line *3*) and $\xi_c(0)$ (line *4*) for the OD single crystal of YBa$_2$Cu$_3$O$_{6.94}$. The values of $\xi_c(0)$ are multiplied by 25. Straight lines are guidance for eye.

It should be emphasized that such an unusual dose dependence is observed only in the results of FLC (Fig. 4) and PG (Fig. 7), which is especially well illustrated by the dose dependences of $T^*$ and $T_{\text{pair}}$ (see Fig. 9, curves 1 and 2). At the same time, oddly enough, $T_c$, resistivity, $d\rho/dT$ and coherence lengths depend linearly on φ, without any features at $\varphi_3$ (Figs. 2 and 9). Thus, comparing the results, we can conclude that *the interaction of radiation defects with normal charge carriers, responsible for resistivity, and with the LPs, responsible for PG, differs significantly.* Figure 9 also shows the dose dependences $\xi_c(0)$ (line 4) and $\xi_{ab}(T_{\text{pair}}) = \xi_{ab}(0)(T_{\text{pair}}/T_c - 1)^{-1/2}$ (line 3), of which the latter is more important since it is related to the dose dependence of $T_{\text{pair}}$ [61]. Strictly speaking, the linear $\xi_{ab}(T_{\text{pair}})$ on φ looks rather unexpected, since the dose dependences of $T_{\text{pair}}$ (curve 2) and $T^*$ (curve 1) have a sharp minimum at $\varphi_3 = 4.3 \cdot 10^{18}$ e/cm$^2$. The absence of features in the dose dependences $\xi_c(0)$ and $\xi_{ab}(T_{\text{pair}})$ (Fig. 9), as well as $\rho(T)$, $d\rho/dT$ and $T_c$ (Fig. 2), allows us to conclude that *there is a certain specific interaction of radiation defects in YBCO with the electronic subsystem,* which leads to the observed linear radiation dependence of these parameters and the nonlinear dependence of $\sigma'(T)$, $\Delta^*(T)$, $T^*$, and $T_{\text{par}}$.

Recall that $\xi_c(0)$ is determined directly from the FLC measurements (refer to Fig. 4). To estimate $\xi_{ab}(T)$, it is assumed that at $\varphi_1 = 0$ we have $\xi_{ab}(0) = \xi_{ab}(T^*) = 10$ Å, which is a typical value of OD YBCO [76, 77], and, depending on φ, it varies with the same law as $\xi_c(0)$. In turn, $\xi_{ab}(T_{\text{pair}})$ at each φ is calculated by the formula $\xi_{ab}(T_{\text{pair}}) = \xi_{ab}(0)(T_{\text{pair}}/T_c - 1)^{-1/2}$, with the corresponding value $\xi_{ab}(0)$, as mentioned above. It should also be kept in mind that $T_{\text{pair}}$ separates both BEC and BCS regimes [12, 28, 61], as mentioned above. It is noteworthy that in well-structured YBCO films with different charge carrier densities $n_f$, as $T_c$ decreases from 87.4 K to 55.4 K, $T_{\text{pair}} = (133 \pm 1)$ K and $\xi_{ab}(T_{\text{pair}}) = (18 \pm 0.5)$ Å remain constant, and do not depend on $n_f$, i.e., on the oxygen content in the samples [61]. In other words, with a change in the oxygen content, the temperature range in which the LPs are transformed from SBBs to fluctuating CPs does not depend on $n_f$. A different picture is observed during irradiation. As can be seen from Table 2, in an unirradiated single crystal, $\xi_{ab}(T_{\text{pair}}) \sim 18$ Å, that is, the same as in the defect free F1 film. However, as the radiation dose increases, $\xi_{ab}(T_{\text{pair}})$ almost doubles (Table 2). In this case, as mentioned above, $\xi_{ab}(T_{\text{pair}})$ is a linear function of φ, regardless of how $T_{\text{pair}}$ changes (Fig. 9). Thus, it can be concluded that the observed unusual dose dependences of the FLC and PG *are in no way related simply to a decrease in the oxygen content* in sample, but *are due to the specific interplay of radiation defects with the internal defect structure of the crystal.*

Figure 10 shows the leftmost points $\Delta^*(T)$ at each radiation dose φ in the range $T_c^{mf} < T < T_{01}$, which should confirm the above conclusions. For a better understanding of the evolution of $\Delta^*(T)$ near $T_c$, the inset to Fig. 7, as an example, shows in detail the shape of $\Delta^*(T)$ at $\varphi_1 = 0$. Interestingly, such a shape of $\Delta^*(T)$ with a pronounced minimum at $T_{01}$, a maximum at $T \leq T_0$ and a small minimum at $T_G$, is characteristic for all without exception HTSCs, including FeAs-based [63] and FeSe [58] superconductors.

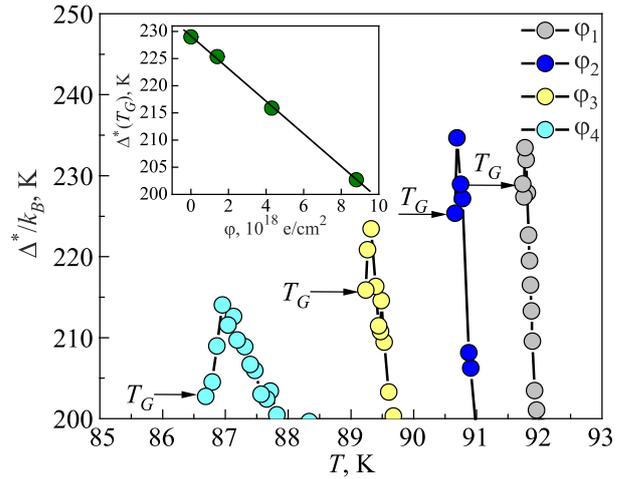

*Fig. 10.* (Color online) The same temperature dependences of the pseudogap $\Delta^*(T)$, as in Fig. 7 at different radiation doses: $\varphi_1 = 0$ (gray dots), $\varphi_2 = 1.4 \cdot 10^{18}$ e/cm$^2$ (blue dots), $\varphi_3 = 4.3 \cdot 10^{18}$ e/cm$^2$ (yellow dots), and $\varphi_4 = 8.8 \cdot 10^{18}$ e/cm$^2$ (turquoise dots) at $T_c^{mf} < T < T_{01}$. The inset shows $\Delta^*(T_G)$ at different φ. Solid curves are guidance for eye.

As can be seen from Fig. 10, the true values of PG, $\Delta^*(T_G) = \Delta(0)$, gradually decrease with increasing φ. This result seems to be reasonable, since, just as in YBCO films [61] and single crystals without defects [75], the lower $T_c$, the less the true value of PG, $\Delta^*(T_G) = \Delta(0)$ [12, 61]. Moreover, despite the revealed feature in the behavior of FLC (Fig. 4), PG (Fig. 7), $T^*$, and $T_{\text{pair}}$ (Fig. 9) at $\varphi_3 = 4.3 \cdot 10^{18}$ e/cm$^2$, $\Delta^*(T_G)$ decreases linearly with decreasing $T_c$ (see inset in Fig. 10). This also suggests that *radiation defects do not much affect the fluctuating Cooper pairs near $T_c$.*

## Conclusion

The effect of high-energy electrons on the fluctuation conductivity (FLC) and pseudogap (PG) in a YBa$_2$Cu$_3$O$_{7-\delta}$ (YBCO) single crystal with pronounced tween boundaries (TBs) has been studied for the first time. The FLC and PG were derived from resistivity measurements and analyzed within the local pairs model. Irradiation with 2.5 MeV electrons leads to a noticeable displacement of O, Cu, Y, and Ba in the crystal and to the formation of a large number of defects due to the displacement of all atoms.

It was shown that, as in previous studies, $T_c$ decreases linearly with irradiation, while the resistivity increases also almost linearly (Figs. 1 and 2). However, in contrast to previous studies, the slope $d\rho/dT$ of the linear part of $\rho(T)$ in the normal state also increases linearly (Fig. 1 and inset in Fig. 2). The incident electrons with the high energy of 2.5 MeV, used in our experiment, with a high probability can remove some of the oxygen atoms from the CuO$_2$ planes of YBCO, thus reducing the density of charge carriers $n_f$ but increasing their scattering rate, especially at high temperatures. This should result in the observed decrease of $T_c$ and an increase of $\rho(T)$ and $d\rho/dT$. At the same time, a sharp unexpected decrease in $T^*$ from 193 K to 136.4 K at $\varphi_3 = 4.3 \cdot 10^{18}$ e/cm$^2$ is observed (Fig. 1 and Table 3). We believe that this is due to a specific interplay between radiation defects and twins and TBs that are initially present in the crystal (refer to the text). The radiation defects should act on the TBs, which will lead to the elimination of the influence of twins and TBs and the equalization of the charge distribution over the sample. This is what leads to a rather peculiar temperature dependence of both FLC (Fig. 4) and PG (Fig. 7) observed in our experiments.

Figure 4 shows that at all $\varphi$ the agreement of the FLC data with the AL and MT theories is very good. This fact suggests that above $T_c$ up to $T_{01}$ the fluctuating Cooper pairs behave like the SC but non-coherent pairs. However, the evolution of FLC with increasing $\varphi$ turned out to be rather unusual. At $\varphi_1 = 0$, above the crossover temperature $T_0$ the deviation of 2D–MT fluctuations over 3D–AL (denoted as $\Delta \ln \sigma'$ in Fig. 4) is quite large: $\Delta \ln \sigma' = 0.68$. This shape of FLC dependence is typical for the single crystal with TBs. At relatively low $\varphi_2 = 1.4 \cdot 10^{18}$ e/cm$^2$, the deviation of fluctuations of the 2D–MT over 3D–AL increases, and $\Delta \ln \sigma'$ takes the greatest value 0.89 (Table 2) suggesting the enhancement of radiation defects in the sample. But the influence of TBs is also still large.

Unexpectedly, with a further increase in the dose to $\varphi_3 = 4.3 \cdot 10^{18}$ e/cm$^2$ both the absolute value of FLC and the deviation $\Delta \ln \sigma' = 0.40$ sharply decrease. Surprisingly, with a further increase in the dose more than twofold to $\varphi = 8.8 \cdot 10^{18}$ e/cm$^2$, the absolute value of FLC is noticeably restored, and the deviation $\Delta \ln \sigma' = 0.21$ takes the smallest value. In fact, the dependence of $\ln \sigma'$ on $\ln \varepsilon$ takes the shape typical of a well-structured YBCO film with a close $T_c = 87.4$ K, denoted in Table 3 as sample F1. Moreover, as with the F1 film, the scaling factor $C_{3D} = 1$ in the Eq. (2) indicates complete agreement between the data and the AL theory. Thus, it seems that when irradiated with high-energy electrons with the highest dose $\varphi$, our *single crystal behaves like a YBCO sample, but without defects.* The finding allowed us to conclude that *the radiation defects completely equalized the distribution of charge, that is, virtually eliminated the effect of twins and twin boundaries in the sample.* In fact, the *greater the disorder caused by defects, the more isotropic the sample.*

The conclusion was completely confirmed by the results of the PG analysis shown in Fig. 7. The evolution of the PG, $\Delta^*(T)$, with increasing $\varphi$ turned out to be even more peculiar than the corresponding dependence of the FLC. With an increase in $\varphi$ the shape of $\Delta^*(T)$ changed from a shape typical for optimally doped OD single crystals containing TBs to a shape characteristic of defect-free YBCO films and annealed single crystals. A certain role in this can be played by specific mechanisms of quasiparticle scattering associated with the presence of kinematic anisotropy in the system [78–81]. Thus, indeed, when irradiated with high-energy electrons with the highest dose $\varphi$, our *single crystal behaves like a YBCO sample without defects.*

It should be also emphasized that such an unusual dose dependence is observed only in the FLC (Fig. 4) and in the PG results (Fig. 7), which is especially well illustrated by the dose dependences $T^*$ and $T_{pair}$ (Fig. 9, curves *1* and *2*). At the same time, oddly enough, $T_c$, resistivity, and $d\rho/dT$ linearly depend on $\varphi$, without showing any specific features at $\varphi_3$ (Fig. 2). Thus, comparing the results, one can draw another important conclusion that *the interaction of radiation defects with normal charge carriers responsible for the resistivity and with LPs differ significantly.* In addition, the true values of PG, $\Delta^*(T_G) = \Delta(0)$ also linearly decrease with increasing $\varphi$ (inset in Fig. 10), which allows us to draw the following conclusion: *radiation defects do not strongly affect the fluctuating Cooper pairs near $T_c$.* The revealed results seem to be well reliable since they distinct observed in both FLC and PG analysis. However, in order to clarify the details, it would be desirable to conduct additional studies of the effect of irradiation on YBCO single crystals with high-energy electrons, but with a smaller step in $\varphi$.

## Acknowledgments

We acknowledge support from the National Academy of Sciences of Ukraine through Young Scientists Grant No. 1/N-2021 (L.V.O. and E.V.P.). This research was funded in part by the Science Committee of the Ministry of Education and Science of the Republic of Kazakhstan (Grant No. AP08052562) and by the Ministry of Innovative Development of the Republic of Uzbekistan (Grant No. Ф-ФА-2021–433). A.L.S. also thanks the Division of Low Temperatures and Superconductivity, INTiBS Wroclaw, Poland, for their hospitality.

---

---

Вплив електронного опромінення на флуктуаційну провідність та псевдощілину в монокристалах $YBa_2Cu_3O_{7-δ}$

A. L. Solovjov, L. V. Omelchenko, E. V. Petrenko, G. Ya. Khadzhai, D. M. Sergeyev, A. Chroneos, R. V. Vovk

Вивчено вплив електронного опромінення з енергією 2,5 МеВ на температурні залежності питомого опору $\rho(T)$ оптимально допованого монокристала $YBa_2Cu_3O_{7-δ}$. Температурні залежності флуктуаційної провідності $\sigma'(T)$ та псевдощілини $\Delta^*(T)$ від дози опромінення $\varphi$ розраховано в межах моделі локальних пар. Показано, що зі збільшенням $\varphi$ значення $\rho(300 K)$ збільшується лінійно, тоді як $T_c$ лінійно зменшується. В той самий час значення $\rho(100 K)$ зростає нелінійно, демонструючи особливість при $\varphi_3 = 4, 3 \cdot 10^{18}$ е/см², що також спостерігається на низці інших залежних від дози параметрів. Незалежно від дози опромінення в інтервалі температур від $T_c$ до $T_{01}$ $\sigma'(T)$ підпорядковується класичним флуктуацій-ним теоріям Асламазова–Ларкіна (3D–АЛ) і Макі–Томпсона (2D–МТ), демонструючи 3D–2D кросовер при підвищенні температури. Температура кросовера $T_0$ дає змогу визначити довжину когерентності вздовж осі $c$ $\xi_c(0)$, яка під впливом опромінення збільшується в $\sim$ 3 рази. Крім того, помітно збільшується інтервал надпровідних флуктуацій вище $T_c$. При $\varphi_1 = 0$ спостерігається залежність $\Delta^*(T)$, типова для монокристалів, що містять яскраво виражені границі двійників, з максимумом при $T_{pair} \sim 120$ K та чітким мінімумом при $T = T_{01}$. Вперше встановлено, що при $\varphi_3 = 4{,}3 \cdot 10^{18}$ е/см$^2$ форма $\Delta^*(T)$ сильно змінюється та стає такою ж, як і в оптимально допованих монокристалах YBa$_2$Cu$_3$O$_{7-\delta}$ з дуже низькою температурою відкриття псевдощілини $T^*$ й помітно зменшеною $T_{pair}$, тоді як на залежності $T_c(\varphi)$ не спостерігається жодних особливостей. Зі збільшенням дози опромінення до $\varphi_4 = 8{,}8 \cdot 10^{18}$ е/см$^2$ форма $\Delta^*(T)$ відновлюється і стає такою ж, як у добре структурованих плівках YBa$_2$Cu$_3$O$_{7-\delta}$ та бездвійникових монокристалах. В цьому випадку $T_{pair}$ та $T^*$ помітно зростають.